# Neutron lifetime splitting in beta-decay

V.V. Vasiliev

NRC "Kurchatov institute" – ITEP

25, Bol. Cheremushkinskaya St., Moscow, 117218, Russian Federation

e-mail basil_v@itep.ru

Abstract

The author considers a hypothesis of neutron lifetime splitting in beta-decay and shows that the beta-decay of neutrons could be described by the triad of lifetimes $\tau_{Left}, \tau_{Mean}, \tau_{Right}$. The lifetime $\tau_{Left}$ is the lifetime of $L$-neutrons emitting electrons against the neutron spin direction ($L$-type neutron decay). The lifetime $\tau_{Right}$ is the lifetime of $R$-neutrons emitting electrons in the direction of the neutron spin ($R$-type neutron decay). The lifetime $\tau_{Mean}$ is the arithmetic average of $\tau_{Left}, \tau_{Right}$ or the mean neutron lifetime. While using the parameters of electron-spin asymmetry of neutron decay and the results for determining the mean neutron lifetime, the performed numerical estimates gave the numerical values of the triad $\tau_{Left}, \tau_{Mean}, \tau_{Right}$ as 813 s, 900 s and 987 s respectively. In addition to the estimates, the lifetimes of the triad are determined from experimental data applying the decay scale tuning method proposed by the author. The experimental values of the triad coincided with the estimates with high accuracy. The weighted average neutron lifetime $\tau_W$ determined from the experimental neutron lifetimes $\tau_{Left}$ and $\tau_{Right}$ is equal to $\tau_W = 883.33 \pm 0.02\,\text{s}$, and is in good agreement with the values of the neutron lifetime obtained by the other methods.

**Introduction.** Experiments on determining the neutron lifetime continue for more than seventy years. Two main methods have been used to determine the neutron lifetime in the past thirty years. The first is the beam method, where the counting rate of protons or electrons from a well-defined volume of a neutron beam of a reactor is measured. The most accurate result of the beam method was obtained by detecting protons in 2013 [1], and is equal to $\tau_n = 887.7 \pm 2.2$ s. The second method is a "bottle method" based on storing ultracold neutrons in a trap until beta-decay. In 2018, this method yielded the neutron lifetime equal to $881.5 \pm 0.9$ s [2]. The previous result of the same group in 2008 [3] was also distinguished by the high accuracy and was $\tau_n = 878.5 \pm 0.8$ s. The result of the V.I. Morozov group [4] obtained by the same bottle method is $880.2 \pm 1.2$ s. The most accurate result of 2014 for measuring the neutron lifetime during storage of ultracold neutrons in magnetic neutron traps amounted to $878.3 \pm 1.2$ s [5]. The problem of systematic errors for experiments using neutron storage is in the accuracy of accounting for neutron escape channels in addition to beta-decay. Estimates are model depending and lead to the shift of estimates beyond the limits of the specified errors.

The most straightforward way to measure the neutron lifetime in beta-decay is experiments at neutron beams of reactors, because the beta-decay of the neutron is directly measured. In beam experiments, the differential decay equation is:

$$\frac{dN_d}{dt} = \frac{1}{\tau_n} \cdot \varepsilon_{eD} \cdot \varepsilon_{nD} \cdot \varepsilon_T \cdot N \,, \qquad (1)$$

where $\frac{dN_d}{dt}$ - electron detector counting rate, $\tau_n$ – neutron lifetime, $\varepsilon_{eD}$ - electron detector efficiency, $\varepsilon_{nD}$ - neutron detector efficiency, $\varepsilon_T$ - efficiency of collection and transport of electrons from the decay volume to the detector, $N$ –total number of neutrons at the time interval $dt$ in the beam region with exactly defined boundaries.

The correct determination of the number $N$ of neutrons and the efficiencies $\varepsilon_{eD}, \varepsilon_{nD}, \varepsilon_T$ is the problem. The latest and the most accurate for the beam experiment results with all sources of systematic errors taken into account most thoroughly is the mentioned one of 2013 [1]. The result with apportion of a systematic error was $\tau_n = 887.7 \pm 1.2\ (stat)\ \pm 1.9\ (syst)\ s$ and based on the correction of experimental data obtained by the authors in 2005 and then published as equal to $\tau_n = 886.3 \pm 1.2(\text{stat}) \pm 3.2(\text{syst})$ [6].

The systematic error of beam experiments includes the accuracy of the cross section for neutron interaction with the neutron detector substance. The authors of [1], [6] directly indicate that with an increase in the accuracy of determining the cross sections for the interaction of neutrons with detector materials, the value of the neutron lifetime will also be recalculated.

Long-term activity in the two discussed directions - beam measurements and bottle measurements - gives, however, significantly different results. The difference is more than 6 seconds, multiply more than the experimental errors indicated by the authors of the works mentioned above.

In addition to the mentioned papers, in 2003 the paper [7] was published with the value for the neutron lifetime $900.00 \pm 0.15$ s. The striking difference between this result with its high accuracy and the results of the predecessors [8] did not contribute to its recognition.

The main goal of this work is to eliminate the sources of systematic errors inherent in the beam method and to take into account the asymmetry of the neutron decay, i.e. the presence in the neutron beta-decay of two frequencies of decay electron generation.

**1. Variation method of the decay scale tuning.** The proposed method [9] is the improving development of the beam method of measuring the neutron lifetime. The novelty of the proposed method is a stepwise change of the number of neutrons in the region controlled by the electron detector. In other words, the goal is to change the so-called neutron numbers, i.e. numbers of neutrons, the decay of which is visible to the electron detector. The important condition is an accurate measurement of the electron-counting rate at each step of variation by repeating each step multiple times. In this case, the problem of measuring the number of neutrons in a beam at each step of the variation is not set.

The system of differential equations of the type (1) for $k$ steps of neutron numbers is:

$$\tau(R_i \pm \sigma_i) \approx \frac{1}{\mu} m_i. \qquad (2)$$

Here $i$ is the step number, $N_i^d \approx \frac{1}{\mu} \cdot m_i$ - are neutron numbers seen for the electron detector, i.e. *numbers of neutrons in the view field of the electron detector*. The parameter $\mu$ is called the scale factor or $\mu$-factor. The neutron numbers are represented here as a limited sample of the certain arithmetic progression members corresponding to the resulting count rates.

Below the simplified notation $N_i^d \equiv N_i$ will be used.

Therefore, the idea is to choose the best decay scale, i.e. $\mu$-factor, and to describe the resulting set of counting rates and their errors by the arithmetic progression of neutron numbers using the least squares method. The method is called "the method of decay scale tuning" (*DST*-method).

For an arbitrary value of the μ-factor, the estimate $\aleph_i$ of neutron numbers $N_i$ for the trial lifetime τ, using the operator "round[*C, p*]" to round *C* to the nearest digit of order *p*, is the following:

$$\aleph_i = round\left[\frac{round\left[\mu \cdot \tau \cdot R_i, \, 0\right]}{\mu}, \, p\right]. \qquad (3)$$

The following error functional is constructed:

$$F_{\mu, p}(\tau) = \sum_{i=1}^{k} \frac{(R_i - \frac{1}{\tau} \cdot \aleph_i (p, \mu, \tau))^2}{\sigma_i^2}. \qquad (4)$$

Note the dependence of the functional (4) on the lifetime due to the operator (3) has the form of a kind of jump function. The term "jump function" is borrowed from the book [10]. The difference is the presence of a jump in the periodic or quasiperiodic function (4). The depth of the jump (and the "contrast" of the graphical representation) of the dependence (4) is maximal at the minimal *p*.

The neutron lifetime $\tau_0$ is determined from the condition of the minimum of the error functional:

$$\left.\frac{dF_{\mu_0, p}(\tau)}{d\tau}\right|_{\tau=\tau_0} = 0. \qquad (5)$$

The solution corresponds to that pair $(\tau_0, \mu_0)$, where condition (5) is implemented with minimum values of μ and the parameter *p* ensuring that the minimum of the functional meets either the number of degrees of freedom ν associated with the number of measurements of counting rates, or the condition of reducing the functional to the value $\chi^2 / \nu$.

A neutron decay nature has a certain specificity associated with the asymmetry of the decay. It is important for the right application of the *DST*-method to the neutron decay.

## 2. A neutron duplicity and lifetime.
The phenomenon of electron-spin asymmetry of the neutron beta-decay $n \rightarrow pe^- \tilde{\nu}$ is well known. The asymmetry means that neutron decays with electron ejection along with the neutron spin direction occur less frequently than neutron decays with electron ejection against the neutron spin direction. The well-known Jackson-Treiman-Wilde formula [11] for the neutron decay probability for the case of registration of electrons only is

$$dW\left(E_e, \Omega_e\right) = W_0 dE_e d\Omega_e \{1 + A \cdot \frac{\mathrm{v}_e}{c} \cdot \cos\Theta_e\},\qquad(6)$$

where is the coefficient of the neutron spin-electron correlation, $\Theta_e$ - electron emittance angle relative to the neutron spin direction, $\frac{\mathrm{v}_e}{c}$ - electron helicity, $W_0$ – constant. From numerous experiments [12] the coefficient $A$ is equal to $A = $ - 0.1173 ± 0.0013. Thus, the decays of neutrons with electron emitted along with the neutron spin and decays emitting electron against the neutron spin differ qualitatively and quantitatively.

Let the observer always look at the neutron so that the neutron spin vector is directed from the left to the right. Then, from the view point of the observer, the set of all neutrons consists of a subset of $L$-neutrons decaying with the emitting of electrons against the neutron spin direction (*Left*) and subset of $R$-neutrons decaying with electron emitted along with the neutron spin direction (*Right*). Hereby, following from the magnitude and sign of the correlation coefficient $A$, the channel $L$ with a negative projection of the electron momentum on the neutron spin takes place about 20% more often than the channel $R$ with a positive momentum direction. Hence during the decay of neutrons two decay frequencies should be observed, corresponding to the decays of $L$-neutrons and $R$-neutrons. Thus, the decay parameters on these neutron subsets ($L$ and $R$) could be defined separately.

The decay constant $\lambda_L$ on the subset of $L$-neutrons as the specific decay frequency of neutrons of this subset is

$$\lambda_L = \frac{1}{N_L} \cdot \frac{dN_L}{dt}.$$

For the constant $\lambda_R$ decay on the subset of $R$-neutrons is

$$\lambda_R = \frac{1}{N_R} \cdot \frac{dN_R}{dt}.$$

The values $\lambda_L$ and $\lambda_R$ could be written as

$$\lambda_L = \lambda_0(1+\Delta),\qquad(7\text{-}1)$$

$$\lambda_R = \lambda_0(1-\Delta).\qquad(7\text{-}2)$$

Here $\lambda_0$ is a mean value of the constants $\lambda_L$ и $\lambda_R$, $\Delta = A \cdot \frac{\overline{\mathrm{v}}_e}{c}$ is an integral asymmetry parameter, $A-$ the coefficient of spin-electron correlation, $\frac{\overline{\mathrm{v}}_e}{c}$ – the mean helicity of electrons.

The total number of neutrons $N_T$ is equal to the sum of the quantities of $L$-neutrons and $R$-neutrons. The total decay constant $\lambda_W$ is the specific decay frequency on the total number $N_T$ of neutrons and must be defined as $\lambda_W = \dfrac{1}{N_T} \cdot \dfrac{dN_T}{dt}$, where $N_T = N_L + N_R$. After substitution, the result for the total decay constant is the formula of weighted average for the total decay constant and exactly corresponds to the classic probability theory: $\lambda_W = \lambda_L \cdot \dfrac{N_L}{N_L + N_R} + \lambda_R \cdot \dfrac{N_R}{N_L + N_R}$ , or by introducing weights of *L*-neutrons and *R*-neutrons: $\lambda_W = \lambda_L \cdot W_L + \lambda_R \cdot W_R$. The channel weight is equal to the ratio of the specific (reduced) channel decay frequency to the total sum of all (here *L* and *R* channels) specific decay frequencies:

$$W_L = \frac{\lambda_L}{\lambda_L + \lambda_R}, \qquad (8\text{-}1)$$

$$W_R = \frac{\lambda_R}{\lambda_L + \lambda_R} . \qquad (8\text{-}2)$$

It is important to emphasize here that the total decay constant $\lambda_W$ is a weighted average of the decay constants $\lambda_L$ and $\lambda_R$ instead of a simple sum of these constants.

It follows from (7) and (8) that the weighted decay constant $\lambda_W$ is related to $\lambda_0$ as

$$\lambda_W = \lambda_0 \cdot \left(1 + \Delta^2\right). \qquad (9)$$

Determining the lifetime of *L*-neutrons through the corresponding decay constant $\tau_{Left} = \dfrac{1}{\lambda_L}$ and the lifetime of *R*-neutrons as $\tau_{Right} = \dfrac{1}{\lambda_R}$, the relations for the lifetimes of *L*-neutrons and *R*-neutrons in dependence on $\Delta$ are

$$\tau_{Left} = \tau_{Mean}(1 - \Delta) , \qquad (10\text{-}1)$$

$$\tau_{Right} = \tau_{Mean}(1 + \Delta) . \qquad (10\text{-}2)$$

The mean neutron lifetime $\tau_{Mean}$ is

$$\tau_{Mean} = \frac{1}{\lambda_0} \cdot \frac{1}{1 - \Delta^2} . \qquad (11)$$

Similarly, after (9) the weighted average lifetime is

$$\tau_W = \tau_{Mean} \cdot \left(1 - \frac{2\Delta^2}{1 + \Delta^2}\right) \qquad (12)$$

Thus, the parameter $\Delta$ may be called as a lifetime asymmetry.

Therefore, the description of the set of electron counting rates in a wide range of values requires the introduction of a neutron lifetime doublet $\tau_{Left}$ and $\tau_{Right}$ because of the presence of two different decay frequencies.

These elementary expressions (9), (10), (11) and (12) provide condition for a numerical estimation of the expected observed values $\tau_{Left}, \tau_{Right}$. The experimental value of $A$ was mentioned above. In addition, the value $\dfrac{\overline{v}_e}{c}$ was estimated as $\dfrac{\overline{v}_e}{c} = 0.824$ from the decay electron spectrum measured in [13]. The value $\tau_{Mean} = 900$ s was taken from the paper [7]. As a result, for $\tau_{Left}, \tau_{Right}$ the following numbers were received: $\tau_{Left} = 813$ s, $\tau_{Right} = 987$ s.

In order to verify these estimates by using the developed *DST*-method, it remains to choose from existing arrays of experimental data the set of electron counting rates from the decay of neutrons. These counting rates should meet the requirements of a variation of initial number of neutrons.

**3. Experimental material.** The experimental material obtained by studying the background in the ITEP experiment on the magnetic storage of ultracold neutrons (UCN) turned out to be quite suitable for the implementation of the proposed method in practice.

In the last series of the ITEP experiment on neutron storage a cycle "inlet-hold-drain to the detector" was performed under special conditions. A UCN absorber film was placed at half-height of the magnetic trap. This feature led to a noticeable reduction in the number of neutrons accumulated in the trap during the inlet time and made it possible to determine the fluctuations of the background more accurately. Only the data that described the counting of electrons flowing from the trap to the detector and forming the background of the experiment were accumulated for processing in this paper. The source of these electrons was the neutron background from the reactor penetrating the walls of the trap [7]. Those background neutrons were thermal, intermediate and fast neutrons from the vertical and horizontal channels of the reactor. The number of background neutrons significantly exceeded the number of UCNs in the trap, especially for the implemented version with the UCN absorber placed in the trap. The data on background measurements in readout intervals were processed and shown in Fig.1a, Fig.1b and Fig.1c. The values of counting rates are arranged in ascending order. The counting rates were defined as average values over readout intervals per storage cycle. The total number of 152 measurements during processing was divided into two sets. There are seventy-one values in the first series ("Series 71", S-71) (Fig. 1a) and eighty-one values in the second series ("series 81", S-81) (Fig. 1b and Fig. 1c).

The long noctidial (more than one hundred days) experiment took place with changes in the set of background sources, since the number of operating reactor channels varied depending on the time of day and also depended on the day of the week, week in month, and month itself. For this reason, the background level varied discretely in accordance with the cyclically varying sets of operating experimental reactor channels that served as sources of the neutron background. Therefore, there are several steps in the figures of counting rates corresponding to the stable combinations of background sources. Switching off the magnetic field of the shutter during the transition from the hold mode to the drain mode led to the leakage of not only the UCN to the detector, which leaked out in a few seconds, but also of accumulated low-energy electrons in the trap. The electrons leaked from the trap much slower than neutrons due to complex quasi-trochoidal trajectories in a non-uniform magnetic field, providing a low drift velocity. On the graph (Fig. 1c), the high levels of count rates correspond to the accumulated electrons. The details and scheme of the experiment were described earlier [7] in detail.

**4. Neutron lifetimes determination.** The expected existence of the triad of lifetimes $\tau_{Left}, \tau_{Mean}, \tau_{Right}$, (10-1), (10-2) and (11), with an appropriate choice of the scale factor could reveal itself in a characteristic dependence of the error functional (4) on the trial lifetime. Since physically *L*-decays and *R*-decays are similar, the picture (form) of the functional in some neighborhood of $\tau_{Left}$ -point will be similar to the picture (form) of the functional in the same neighborhood of the $\tau_{Right}$ − point. A characteristic picture of the error functional around the local center of symmetry forms the functional core. The presence of jumps on both sides of the core center gives the core a contrast emphasizing the core symmetry relative to the solution point, i.e. local lifetime of $\tau_{Left}$ or $\tau_{Right}$ type. Since the mean lifetime is only a consequence of averaging of $\tau_{Left}$ and $\tau_{Right}$, then, with an appropriate step of the scale, in the vicinity of the point $\tau_{Mean}$ the functional will have an identical core shape. The starting point for finding the proper scale step is, firstly, the mean lifetime indicated in [7], and secondly, the possibility of reducing the functional to the unity in the surroundings of this value point by selecting the scale step. All three decision points i.e. $\tau_{Left}, \tau_{Mean}, \tau_{Right}$ will be described by similarly shaped functional cores with core centers at the points of local minima close to the unity. Reducing the error functional to the unity by adjusting the decay scale provides conditions to determine the error value for the result directly from the dependence of the functional on $\tau$ in the close vicinity of the solution point.

Figures 2a, 2b and 2c show the results of reducing the functional to the level $\chi^2 = 1$ in the region of the minimum in the indicated assumed centers of symmetry in the following order: $\tau_{Mean}, \tau_{Left}, \tau_{Right}$. For the complete series of S-152, the optimal scale factor $\mu = 138$ was found by selection. In Fig. 2a, near the value 900 s, for the μ-factor $\mu = 138$ and the accuracy parameter $p = 2$, the core of the functional in the center of symmetry is shown.

The core of the functional to the left of the indicated common center similar in configuration to the central core, with its own local center near the lifetime value of 813 s was found (Fig. 2b). The similar core to the right near the point 987 s was found (Fig. 2c). Comparing these cores from the left to the right, one can identify symmetry attributes between the left and the right cores relative to the neutral central core and clear signs of the similarity of the shape of all three cores. The accuracy level $p = 2$ provides a high "contrast" due to the deep drop in jumps and makes it easy to identify cores. The following indications are obvious for the centers of symmetry in the complete series S-152: the left point $\sim 813$ s, the mean point$\sim 900$ s, and the right point$\sim 987$ s.

The analysis of the minima of reduced functionals in the centers of the right and the left cores is shown in Fig. 3a and Fig. 3b. The reduction $\chi^2 / \nu$ to the unit to the right of the common center is already implemented on the accuracy $p = 2$ (Fig. 3a). Moreover, for the scale factor $\mu = 138$, the reduction result is $\chi^2 / \nu = 1.03$, and for $\mu = 151$, the most exact result is $\chi^2 / \nu = 1.01$. The scale factor ensuring the reduction of the functional to the unity at this point corresponds to the number of degrees of freedom for the number of count rate measurements equal to 152. In particular, the value $\mu = 138$ is in the interval ($\nu \pm \sqrt{2\nu}$) for $\nu = 151$. The lifetime for $R$-neutrons and its error clarify themselves by the dependence of the error functional on the lifetime: $\tau_{Right} = 986.97 \pm 0.03$ s, CL=95% (Fig.3a).

Fig. 3b shows the results of the separate error functional reduction to $\chi^2 / \nu = 1$ for the series S-71 and S-81 in the left core of the functional. The dependence of the averaged functional on the lifetime gives for $\tau_{Left}$ the following result $\tau_{Left} = (162601 \pm 4) \cdot 5 \cdot 10^{-3}$ s. With rounding the error to the first significant digit after the decimal point, the total error of about 0.02 s was obtained. The full result of the $L$-neutron lifetime is $\tau_{Left} = 813.005 \pm 0.02$ s. For the mean neutron lifetime, the result value is $\tau_{Mean} = 899.99 \pm 0.02$ s. This result confirms the result of [7] with higher accuracy.

## 5. Analytical estimation for the neutron lifetime and its error.

Analysis of expressions (4) and (5) gave formulas for the numerical estimate of the neutron lifetime and its error for the optimal set of neutron numbers. After introducing the notations: $A = \sum_i^k \frac{\aleph_i^2}{\sigma_i^2}$, $B = \sum_i^k \frac{R_i \cdot \aleph_i}{\sigma_i^2}$ the formulas for the lifetime and its error are:

$$\tau = \frac{A}{B}, \tag{13}$$

$$\delta\tau = \frac{A^{3/2}}{B^2}. \tag{14}$$

All errors are calculated by the error transfer rule.

The asymmetry of the neutron lifetimes is determined from the lifetimes of $L$-neutrons and $R$-neutrons by the ratio:

$$\Delta = \frac{\tau_{Right} - \tau_{Left}}{\tau_{Right} + \tau_{Left}}. \tag{15}$$

After using formulas (8) and (9) and the notation $\eta = {\tau_{Right}}/{\tau_{Right}}$, the short expression to calculate the weighted average neutron lifetime is:

$$\tau_W = \tau_{Left} \cdot \frac{1+\eta}{1+\eta^2}. \tag{16}$$

The neutron lifetime asymmetry $\Delta$ (15) is equal to $\Delta = (9665 \pm 2) \cdot 10^{-5}$ and coincides with $\frac{1}{\sqrt{107}}$ within the error. This value is in full agreement with the electron-spin correlation coefficient and the average helicity of electrons adopted earlier for the initial estimation of $\Delta$.

For reference, the calculation results are presented below. The formulas (13) and (14) for the left core of the functional in the case of the S-71 series with a scale factor of $\mu = 69$, accuracy order $p = 4$, were used (Fig. 3b). After applying the *DST*-method to the data array shown in Fig. 1-1, with using formulas (3), (4), (5), the results for $A$ and $B$ are the following: $A = 2.75814 \cdot 10^{14}; B = 3.39253 \cdot 10^{11}$. From these numbers for the lifetime and its error the result $\tau_{Left} \pm \sigma = 813.004 \pm 0.01$ s, 65% CL was received. Hence, the full compliance with the result mentioned above for the $L$-neutron lifetime $\tau_{Left}$ is confirmed.

In addition, eventually, from the results of section 4, using formula (16), the calculated from optimal neutron numbers weighted average neutron lifetime is equal to $\tau_W \pm 2\sigma = 883.33 \pm 0.02$ s, 95% CL.

**6. Results.** Thus, the use of the decay scale tuning method allows obtaining the following results:

- The neutron beta-decay can be described by the triad of neutron lifetimes: $\tau_{Left}, \tau_{Right}, \tau_{Mean}$, i.e. the lifetime of *L*-neutrons emitting an electron against the direction of its own spin when decaying, the lifetime of *R*-neutrons emitting an electron along with the spin direction, and the mean neutron lifetime;

- The numerical values for the new physical quantities from experimental data were determined: the lifetime of *L*-neutrons $\tau_{Left} = 813.005 \pm 0.02$ s and the lifetime of *R*-neutrons $\tau_{Right} = 986,97 \pm 0.03$ s;

- The mean value of the neutron lifetime $\tau_{Mean} = 899,99 \pm 0,02$ s was determined;

- The observed neutron lifetime as $\tau_W = 883,33 \pm 0.02$ s was determined.

- The value of the asymmetry of the neutron lifetimes $\Delta = (9665 \pm 2) \cdot 10^{-5}$ was obtained.

The proposed *Decay Scale Tuning* method enables a performing of precision experiments to study decays at low counting rates.

**7. Conclusions.** It is shown that the neutron beta-decay is a two-channel process. The concept of *L*-neutrons emitting electrons against the neutron spin direction and *R*-neutrons emitting electrons along with the direction of the neutron spin in the beta-decay is introduced. The *DST*-method for the independent determination of the *L*-neutron lifetime and *R*-neutron lifetime is proposed. The exact values of these quantities with applying the new method to the experimental data were obtained. The result for the weighted value of the neutron lifetime is in a good agreement with the results [1], [2] (Fig.4).

The author obtained the experimental data used in this paper with the ITEP UCN group under the guidance of V.V. Vladimirsky at the Heavy Water Reactor of the ITEP, Moscow, Russia in the last session of the CENTAUR ("КЕНТАВР") set-up. The author is deeply grateful to V.V. Vladimirsky for his support at the stage of work in 2002-2003.

The work on this paper has been carried out over and above the state assignment plan and without any financial support.

**Captions for the drawings**

Fig. 1a. Electron counting rates in order of growth, depending on the number in the S-71 series.

Fig. 1b. Electron counting rates depending on the number (72-140), S-81 series.

Fig. 1c. Electron counting rates depending on the number (141-152), S-81 series.

Fig. 2a. The central core of the error functional $F_{\mu=138,p=2}(\tau)$ for the complete S-152 series.

Fig. 2b. The left core of the error functional $F_{\mu=138,p=2}(\tau)$ for the complete S-152 series.

Fig. 2. The right core of the error functional $F_{\mu=138,p=2}(\tau)$ for the complete S-152 series.

Fig. 3a. The lifetime $\tau_{Right}$ in the right core of the reduced functional $F_{\mu=138,p=2}^{(S-152)}(\tau)$,

$\quad\quad$ **1**-$\mu$=151; **2** - $\mu$=138; **3**- $\mu$=69.

Fig. 3b. The lifetime $\tau_{Left}$ in the reduced functional combination for two series:

$\quad\quad$ **1**- S-71- $F_{\mu=69,p=4}^{(S-71)}(\tau)$; **2**- S-81- $F_{\mu=46,p=3}^{(S-81)}(\tau)$; **3** -$\left(F_{\mu=69,p=4}^{(S-71)}(\tau)+F_{\mu=46,p=3}^{(S-81)}(\tau)\right)/2$.

Fig. 4. Comparison of the most accurate results of measuring the neutron lifetime by the three methods: 1-Beam method, 2-Method of storage of ultracold neutrons, 3-*DST*-method.

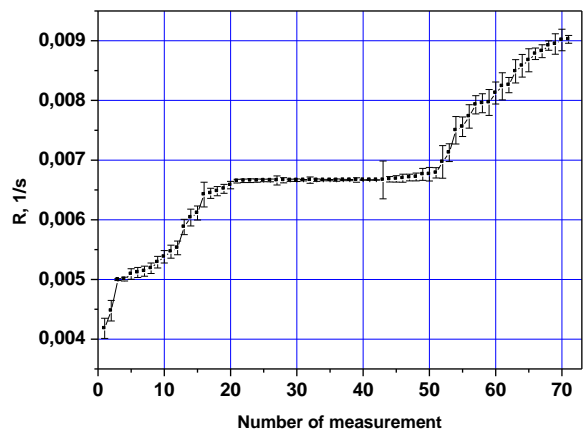

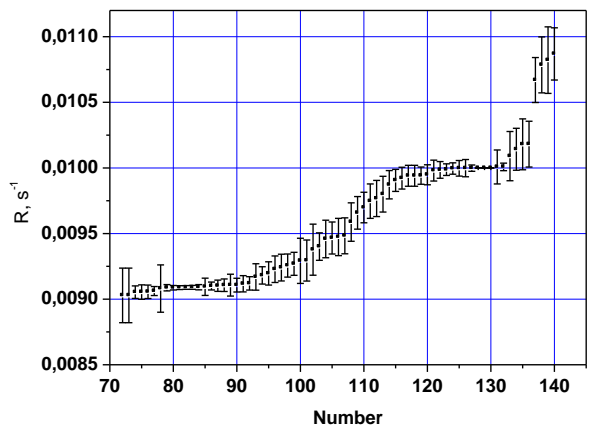

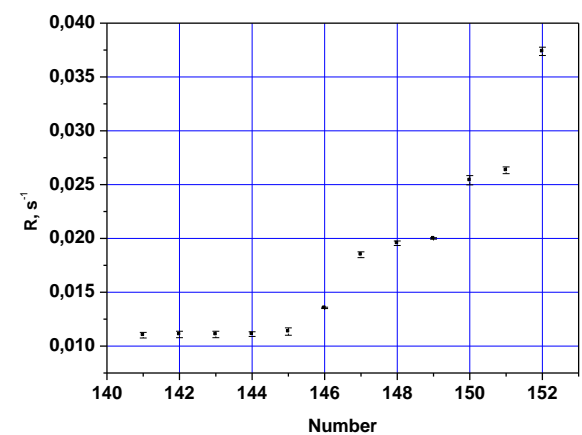

Fig. 1a. Electron counting rates in order of growth, depending on the number in the S-71 series

Fig. 1b. Electron counting rates depending on the number (72-140), S-81 series

Fig. 1c. Electron counting rates depending on the number (141-152), S-81 series

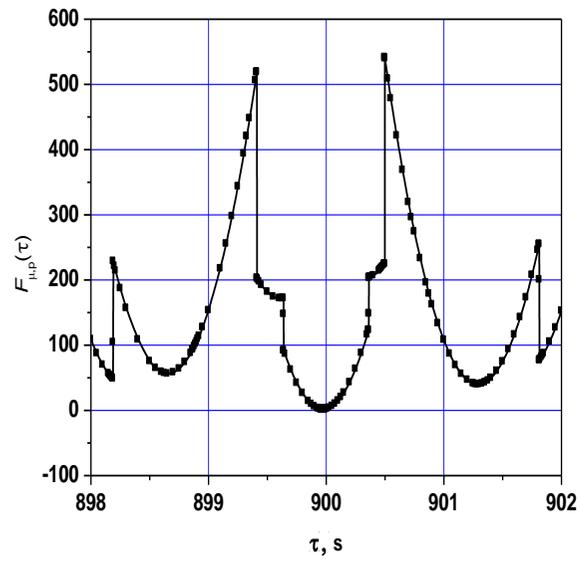

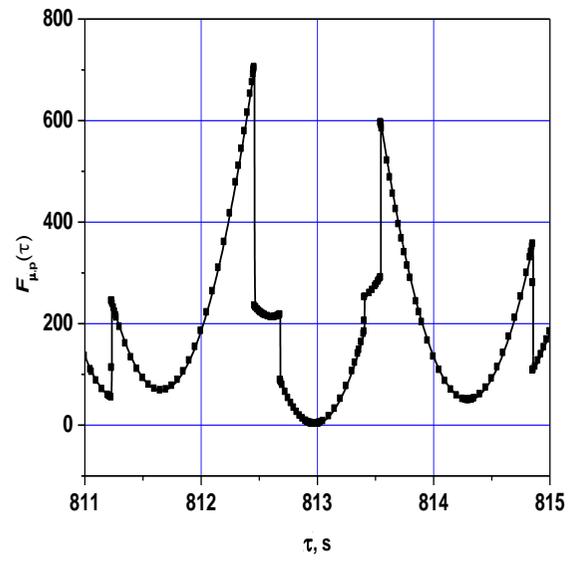

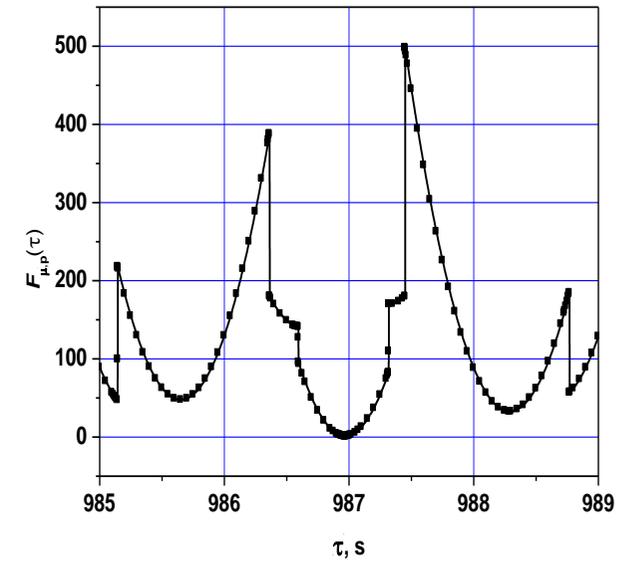

Fig. 2a. The central core of the error functional $F_{\mu=138,p=2}(\tau)$ for the complete S-152 series

Fig. 2b. The left core of the error functional $F_{\mu=138,p=2}(\tau)$ for the complete S-152 series

Fig. 2c. The right core of the error functional $F_{\mu=138,p=2}(\tau)$ for the complete S-152 series

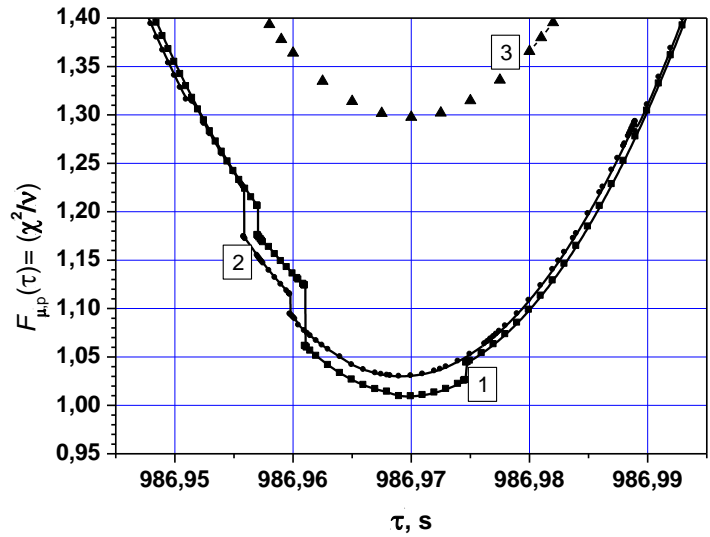

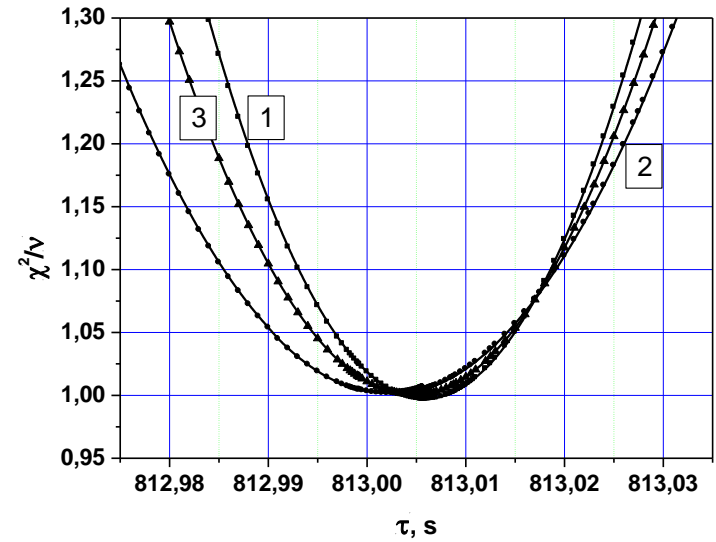

Fig. 3a. The lifetime $\tau_{Right}$ in the right core of the reduced functional $F^{(S-152)}_{\mu=138, p=2}(\tau)$, **1**-$\mu$=151; **2** - $\mu$=138; **3**- $\mu$=69.

Fig. 3b. The lifetime $\tau_{Left}$ in the reduced functional combination for two series: notation: **1**- S-71- $F^{(S-71)}_{\mu=69, p=4}(\tau)$; **2**- S-81- $F^{(S-81)}_{\mu=46, p=3}(\tau)$; **3** -$\left( F^{(S-71)}_{\mu=69, p=4}(\tau) + F^{(S-81)}_{\mu=46, p=3}(\tau) \right) / 2$.

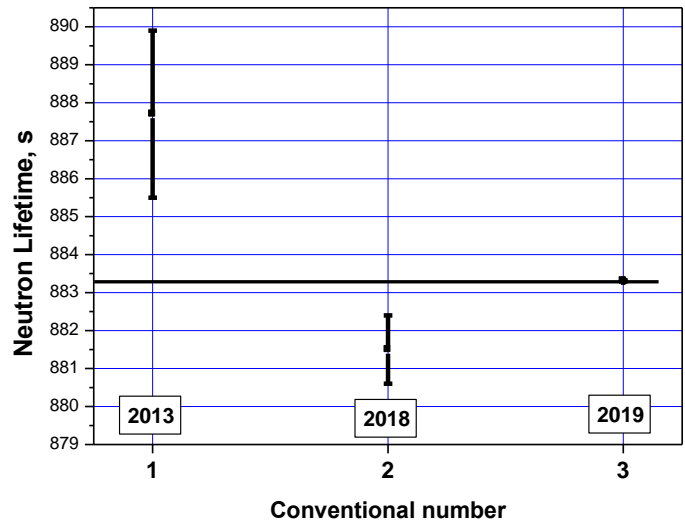

Fig. 4. Comparison of the most accurate results of measuring the neutron lifetime by three methods, 1-Beam method, 2-Method of storage of ultracold neutrons, 3-*DST*-method.